\documentclass[aps,pra,onecolumn]{revtex4-1}

\usepackage[normalem]{ulem}
\usepackage{amsmath,amssymb}
\usepackage{multirow}
\usepackage{xcolor,soul}
\usepackage{srcltx}
\usepackage{hyperref,graphicx}

\begin{document}

\title{Robust light transport in non-Hermitian photonic lattices}

\author{Stefano Longhi}
\email{Corresponding author email: stefano.longhi@polimi.it}
\author{Davide Gatti}
\author{Giuseppe Della Valle}

\vspace{3cm}
\affiliation{Dipartimento di Fisica- Politecnico di Milano and Istituto di Fotonica e Nanotecnologie - Consiglio Nazionale delle Ricerche \ \\ Piazza Leonardo da Vinci, 32, I-20133 Milano, Italy}
\begin{abstract} {\bf 
Combating the effects of disorder on light transport in micro- and nano-integrated photonic devices is of major importance from both fundamental and applied viewpoints.  In ordinary waveguides, imperfections and disorder cause unwanted back-reflections, which hinder large-scale optical integration. Topological photonic structures, a new class of optical systems inspired by quantum Hall effect and topological insulators, can realize robust transport via topologically-protected unidirectional edge modes. Such waveguides are realized by the introduction of synthetic gauge fields for photons in a two-dimensional structure,  which break time reversal symmetry and enable one-way guiding at the edge of the medium. Here we suggest a different route toward robust transport of light in lower-dimensional (1D) photonic lattices,  in which time reversal symmetry is broken because of  the {\it non-Hermitian} nature of transport. While a forward  propagating mode in the lattice is amplified, the corresponding backward propagating mode is damped, thus resulting in an asymmetric transport that is rather insensitive to disorder or imperfections in the structure. Non-Hermitian transport in two lattice models is considered: a tight-binding lattice with an imaginary gauge field (Hatano-Nelson model), and a non-Hermitian driven binary lattice. In the former case transport in spite of disorder is ensured by a mobility edge  that arises because of  a non-Hermitian delocalization transition. The possibility to observe non-Hermitian delocalization induced by a synthetic 'imaginary' gauge field is suggested using an engineered coupled-resonator optical waveguide (CROW) structure.\par}
\end{abstract}
\maketitle

The realization of photonic devices at the micro- and nano-scale capable of guiding light in a controllable way despite the presence of disorder or imperfections is of major importance  for integrated optic applications. Inspired by the quantum Hall effect and the concept of topological insulators in condensed-matter physics, several recent works have proposed and experimentally demonstrated one-way edge propagation for light waves in two-dimensional photonic structures (for a recent review see \cite{r1}).  Topologically protected edge states are one-way guided modes propagating (clockwise or counter-clockwise) along the edge of the sample that cannot be scattered into other states, and are therefore immune to back reflection and localization. The existence of such states requires to break time reversal symmetry of the optical system using magnetic materials, such as in magnetic photonic crystal systems \cite{r2,r3,r3bis},  or by creating synthetic gauge fields for photons. The latter have been demonstrated in a wide variety of optical systems, such as  coupled silicon resonators \cite{r4,r5,r6}, twisted waveguide lattices \cite{r7}, bianisotropic metamaterials \cite{r8}, and lattices with parameters modulated in time \cite{r9}. The possibility to observe and adiabatically transport edge states in quasi-crystals has been reported as well \cite{r10}. \par
 In this work we suggest a different route toward robust propagation in one-dimensional (1D) photonic lattices, in which time reversal symmetry breaking is realized by exploiting non-Hermitian dynamics.  Wave propagation in non-Hermitian (complex) crystals can show a wide variety of interesting effects as compared to ordinary (Hermitian) crystals, such as  double refraction and nonreciprocal diffraction \cite{r11}, unidirectional Bragg scattering and invisibility \cite{r12,r13,r14}, anomalous dynamic localization \cite{r14b}, hyperballistic transport \cite{r15}, and a transition from ballistic to diffusive regimes \cite{r16}. Such effects have been mainly investigated in case of $\mathcal{PT}$ symmetric systems \cite{r17}.  However, in such systems wave transmission turns out to be independent of the incidence side, so $\mathcal{PT}$ structures are unsuited to realize robust one-way transport. Indeed, asymmetric transmission and optical isolation here requires to exploit e.g. nonlinearity in the system \cite{palle1,palle2}. Other possibilities to introduce asymmetric transport is the use of 'imaginary' gauge fields or modulation in time of some parameters of the optical medium. In a pioneering paper, Hatano and Nelson \cite{r18} showed that an 'imaginary' magnetic field in a disordered 1D lattice can induce a delocalization transition, i.e. it can prevent Anderson localization (see also \cite{r19,r20,r21,r22}). In the disordered lattice, delocalized states correspond to complex energies, whereas real energies are associated with Anderson-like localized states.  This is a very interesting result because it shows that mobility is not prevented by disorder in the non-Hermitian 1D lattice with an 'imaginary' gauge field, contrary to a 'real' magnetic field that has no effect on Anderson localization in a 1D chain. Such a result, however,  has been overlooked in the condensed matter physics context, because of the challenging task of synthesizing an 'imaginary' vector potential. In optics, however, this can be realized in coupled microring structures, as we will discuss in our work. We also suggest a more general and physically simple understanding of robust transport in non-Hermitian lattices with asymmetric transmission, the imaginary gauge potential being a special example. 

\section*{Results}
\noindent\textbf{One-way transport in non-Hermitian  photonic lattices: Introductory physical idea.} 
One-way propagation in 2D Hermitian lattices with a synthetic gauge field is based on the existence of topologically-protected edge states. Here we suggest a different  mechanism of one-way robust propagation in a non-Hermitian 1D lattice, in which a {\it preferred} propagation direction arises because of unbalanced amplification/attenuation of waves propagating in opposite directions.
The main physical distinction between topologically protection of edge waveguides in a Hermitian 2D lattice and asymmetric transport in a non-Hermitian lattice can be readily understood in reciprocal space, as schematically illustrated in Fig.1.
In an Hermitian 2D lattice with a synthetic gauge field, the dispersion diagram $E_w=E_w(q)$ of the edge guided mode shows the typical behavior depicted in Fig.1(a). It connects the bulk frequency bands above and below the frequency gap. The unidirectionality of the edge guided mode can be seen from the group velocity $v_g=(dE_w/dq)$  of the waveguide dispersion curve, which has only positive (or only negative) value. In addition, there are no counter-propagating modes at the same frequencies as the one-way edge mode. This enables light to flow around imperfections with perfect transmission. Let us now consider a non-Hermitian 1D lattice with a transmission band described by the energy dispersion curve $E=E(q)$, such that the following condition is satisfied
\begin{equation}
v_g(q) {\rm Im} (E(q)) > 0
\end{equation}
where $q$ is the Bloch wave number, which varies in the interval $ -\pi \leq q < \pi$, and $v_g(q)={\rm Re} (dE/dq)$ is the group velocity [Fig.1(b)]. In this case, a forward   
propagating mode, corresponding to $v_g>0$, is amplified, whereas the corresponding backward propagating mode, with a group velocity $v_g<0$, is damped. This unbalanced amplification/attenuation of waves introduces a preferred direction of propagation, and back scattered  waves arising from disorder or imperfections in the lattice experience attenuation. Thus localization effects are expected to be prevented.
  
\vspace{0.3cm}
\noindent\textbf{Imaginary gauge field and non-Hermitian delocalization.}
The simplest example of a non-Hermitian 1D tight-binding lattice with a dispersion relation satisfying the condition (1) is provided by the Hatano-Nelson model \cite{r19}. Originally introduced to study the motion of magnetic flux lines in disordered type-II superconductors, it describes the hopping dynamics of a quantum particle on a tight-binding ring lattice  threaded by an imaginary magnetic flux. In the absence of disorder, the corresponding Hamiltonian reads  \cite{r19,r20}
\begin{equation}
\hat{H}= \sum_n \kappa \left\{ \exp(-h) | n \rangle \langle n+1| + \exp(h) | n+1 \rangle \langle n | \right\}
\end{equation}
where $\kappa$ is the hopping rate between adjacent sites and the parameter $h$ describes the effect of an imaginary vectorial potential. A simple physical implementation of an imaginary gauge field for photons in a CROW structure will be discussed in details in a following section. Here we note that, for an infinitely-extended lattice, the dispersion relation of the tight-binding lattice band is given by
\begin{equation}
E(q)=2 \kappa \cos(q+ih)= 2 \kappa_1 \cos q -2 i \kappa_2 \sin q 
\end{equation}
  where we have set $\kappa_1=\kappa \cosh h$ and $\kappa_2=\kappa \sinh h$.  The group velocity is given by $v_g={\rm Re}(dE /dq)=-2 \kappa_1 \sin q$. Note that a forward-propagating wave ($-\pi<q<0$, $v_g>0$) is amplified because ${\rm Im}(E(q))>0$, whereas a backward propagating wave ($0<q< \pi$, $v_g<0$) is attenuated because ${\rm Im}(E(q))<0$. This makes wave transport in the lattice highly asymmetric because  backward propagating waves vanish after some propagation distance due to damping. As a result, unidirectional transport is expected to be robust against disorder and lattice imperfections. For disorder in the site energies, the Hamiltonian takes  the form
\begin{equation}
\hat{H}= \sum_n \kappa \left\{ \exp(-h) | n \rangle \langle n+1| + \exp(h) | n+1 \rangle \langle n | \right\}+ \sum_n V_n |n \rangle \langle n |
\end{equation}
  where $V_n$ is a random potential. $V_n$ is assumed independently for each
site, from a distribution with zero mean and variance $u^2$. After setting $| \psi(t) \rangle= \sum_n c_n(t) |n \rangle$, the evolution equations for the amplitude probabilities $c_n(t)$ read explicitly
  \begin{equation} 
  i \frac{dc_n}{dt}=\kappa \exp(-h)c_{n+1}+ \kappa \exp(h) c_{n-1}+V_nc_n.
  \end{equation} 
  For $h=0$, it is known that all the eigenstates of $\hat{H}$ become localized owing to Anderson localization, regardless of the strength $u^2$ of disorder.  However, 
for $h \neq 0$ some delocalized states survive, i.e. a delocalization transition is observed \cite{r18,r19}. Rather generally, the delocalized (localized) states correspond to complex (real) energies, however the precise nature of the non-Hermitian eigenstates is somewhat controversial \cite{r20,r21}. As discussed in previous works, by introducing an imaginary vector potential  a mobility region opens up near the center of the tight-biding energy band, where Anderson localized states show the largest localization length, i.e. they are weakly localized by disorder. Such a phenomenon can be explained as follows \cite{r19}. The eigenstates of the Anderson lattice with the imaginary vector potential can be mapped into the ones of the Hermitian Anderson lattice (i.e. with $h=0$) after the transformation $c_n=a_n \exp(hn)$. Hence, if $a_n$ is a localized state in the Hermitian Anderson lattice with energy $E=E_0$ and localization length $\xi=\xi(E_0)$, it remains a localized state in the non-Hermitian Anderson lattice provided that $2 |h|< 1/ \xi(E_0)$. Such a condition in violated in the range of energies $-E_c < E_0<E_c$ around $E_0=0$, i.e. a mobility region is opened with an edge at $E_c$. The mobility edge $E_c$ can be computed following the analysis of Ref. \cite{r19} and reads explicitly $E_c \simeq (4 \kappa^2 - u^2/2|h|)^{1/2}$. The mobility region opened by the imaginary gauge field ensure particle propagation along the lattice in spite of disorder. This is shown, as an example, in Fig.2. In Figs.2(a) and (b) a semi-infinite lattice is considered, and  the edge site at $n=0$ is occupied at initial time $t=0$. The figure shows the numerically-computed evolution of the wave packet center of mass 
\begin{equation}
\langle n(t) \rangle  =  \frac{\sum_n n |c_n(t)|^2}{ \sum_n |c_n(t)|^2}
\end{equation}
for the Hermitian [Fig.2(a)] and non-Hermitian [Fig.2(b)] lattices.
As in the former case $\langle n (t) \rangle$  clearly ceases to increase as $t$ increases (this is a clear signature of Anderson localization), in the
latter case a secular growth of  $\langle n (t) \rangle$ is observed, indicating the existence of delocalized (mobility) states. The robustness of the transport against disorder in the 
non-Hermitian lattice is also observed when the lattice is initially excited in a single site in the bulk, i.e. far from the lattice boundaries, or by a localized wave packet. As an example, Fig.2(c) showa the numerically-computed dynamics in the disordered lattice with initial excitation of site $n=100$. Since the non-Hermitian dynamics does not conserve the norm, for the sake of clearness in the figure the evolution of the site occupation probabilities for the normalized amplitude probabilities $a_n(t) \equiv c_n(t) / \sqrt{ \sum_{n}|c_n(t)|^2}$ is depicted.  The figure clearly shows that, in spite of disorder, transport is observed in the forward direction. The same scenario is found for  initial wave packet excitations, as discussed in the Methods. In this case, even in the absence of disorder, owing to the dependence of the complex energy dispersion curve $E(q)$ on the Bloch wave number $q$ [Eq.(3)], during the propagation an initial wave packet of the form $c_n(0)=F(n) \exp(iq_0 n)$, with carrier Bloch wave number $q_0$ and slowly-varying amplitude $F(n)$, suffers for reshaping (distortion) effects because of {\it both} group velocity (phase) and amplification (amplitude) dispersion, i.e. because  ${\rm Re}(d^2 E/ dq^2) \neq 0$ and ${\rm Im}(dE/ dq) \neq 0$. For the Hatano-Nelson model, both group velocity and amplitude dispersion effects are minimized at $q_0= \pi/2$, where ${\rm Re}(d^2 E/ dq^2) = {\rm Im}(dE/ dq) = 0$. Examples of wave packet propagation are discussed in the Methods. Finally, it should be pointed out that non-Hermitian transport is robust also against structural imperfections or defects in the lattice. Let us consider, as an example, a lattice with two potential defects at sites $n_0$ and $n_1$, i.e. let us assume in Eq.(4) $V_n=V_0 (\delta_{n,n_0}+\delta_{n,n_1})$, where $V_0$ is the strength of the potential defect. In the Hermitian lattice $(h=0$), a propagating  wave packet undergoes multiple reflections back and forth between the two defects, like in a Fabry-Perot cavity. This yields multiple transmitted wave packets, i.e. echoes of the original wave packet, as illustrated in Fig.3(a). Application of the imaginary gauge field to the lattice ($h \neq 0$) suppresses multiple reflections (echo effects), as shown in Fig.3(b). 
    
\vspace{0.3cm}

\noindent\textbf{Non-Hermitian driven lattice model.}  We conjecture that the absence of Anderson localization and asymmetric transport induced by an imaginary gauge field is a general feature of a non-Hermitian 1D lattice with a dispersion relation of the energy band that satisfies Eq.(1), i.e. that provides amplification (attenuation) for forward (backward) propagating waves. In a 1D lattice with local complex site potentials, transport is always symmetric (reciprocal) \cite{Muga}. However, non-reciprocal (asymmetric) transport can be obtained by introduction of some periodic modulation of system parameters. In this case,  the quasi-energy bands of the driven lattice may satisfy Eq.(1),  i.e. asymmetric transmission can arise. To support our conjecture, let us consider as an example the periodically-driven non-Hermitian binary lattice described by the Hamiltonian
\begin{equation}
\hat{H}= \sum_n \kappa \left\{ | n \rangle \langle n+1| +| n+1 \rangle \langle n | \right\}+\sum_n \left\{ F n +(-1)^n g(t) \right\} |n \rangle \langle n | +\sum_n V_n |n \rangle \langle n |
\end{equation}
where $\kappa$ is the hopping rate between adjacent sites, $F$ is an index gradient along the lattice, $V_n$ is a random potential that accounts for disorder, and
\begin{equation}
g(t)=G_R \cos(\omega t) + i G_I \sin(\omega t)
\end{equation}
is a sinusoidal function of frequency $\omega$ that describes a periodic modulation  of the complex energy sites with opposite sign at alternating sites.  Such a non-Hermitian binary lattice can describe temporal hopping of light in a CROW structure with an impressed static index gradient and with temporal modulation of gain/loss at altering sites in the chain \cite{r22bis}. Also, provided that the temporal dynamics is replaced by spatial propagation, Eq.(7) can describe spatial light transport in an array of circularly-curved evanescently-coupled optical waveguides with alternating gain/loss and index modulation in the guides along the propagation direction $t$ \cite{r15,r23}. The disorder $V_n$ accounts for imperfections in the propagation constants of the various waveguides in the array. Note that the lattice becomes Hermitian in the $G_I=0$ limit. In the absence of disorder, $\hat{H}$ is periodic in time with period $T= 2 \pi / \omega$, and Floquet theory applies. Provided that the resonance condition $F=(M/N) \omega$ is satisfied for some irreducible integers $M$ and $N$,  the lattice sustains two quasi-energy minibands with dispersion relations that satisfy Eq.(1) (see Methods). As an example, Fig.4(a) shows the numerically-computed quasi energy spectrum of the two mini bands for  $\omega / \kappa =1$, $F/ \omega=2$, $G_R / \kappa=4.7$ and $G_I / \kappa=4.26$. The quasi-energy spectrum satisfies the condition (1) and hence forward (backward) waves are amplified (damped) as they propagate in the modulated lattice. The robustness of forward-propagation in the presence of disorder is demonstrated in Figs.4(b) and (c). A semi-infinite lattice in considered,  and excitation of the edge site is accomplished at initial time $t=0$. Figure 4(b) depicts the evolution of the wave packet center of mass $\langle n(t) \rangle$ for the driven lattice in the Hermitian limit $G_I=0$, clearly showing localization in the presence of disorder. Conversely, in the non-Hermitian case propagation is not prevented by the disorder [Fig.4(c)]. A similar scenario is found by considering initial excitation of a site in the bulk.  Robust transport occurs for an initial distribution corresponding to a localized wave packet  as well (see Methods). Interestingly, as compared to the Hatano-Nelson model, in the driven lattice model (5) wave packet  distortion effects arising from group velocity and amplification dispersion are negligible because in this case one has  ${\rm Re}(d^2 E/ dq^2) \simeq 0$ and ${\rm Im}(dE/ dq) \simeq 0$ almost everywhere inside the Brillouin zone [Fig.4(a)].

\vspace{0.3cm}

\noindent\textbf{Realization of an imaginary gauge field in coupled resonator optical waveguides and non-Hermitian delocalization.}  Synthetic gauge fields and robust transport via topologically-protected edge modes in 2D CROW structures have been proposed and experimentally demonstrated in a series of recent works using coupled microring resonators \cite{r4,r5,r6}. Here we show that, by inclusion of engineered gain and loss terms in auxiliary ring resonators, a synthetic 'imaginary' gauge field can be readily implemented in a linear (1D) chain of microring resonators, which can thus provide an accessible optical system for the observation of non-Hermitian delocalization \cite{r18}. A schematic of the 1D chain of coupled ring resonators is shown in Fig.5(a).  The main ring resonators are  indirectly coupled using another set of auxiliary rings which
are designed to be antiresonant to the main ring resonators, i.e., the length of the connecting rings is slightly larger (or smaller) than the main rings so as to acquire an extra $\pi$ phase shift. The auxiliary ring provides amplification in the upper half perimeter, with single-pass amplification $h$, and balanced loss in the lower half perimeter, with single-pass attenuation $-h$. Indicating by $a_n(t)$ the amplitude of the counterclockwise propagating field in the $n$-th ring in the main resonators, with a carrier frequency coincident with one longitudinal ring resonance, coupled-mode equations for the slowly-varying amplitudes $a_n(t)$ can be derived in the mean-field limit after elimination of the field amplitudes in the auxiliary resonators (see Methods). In the absence of disorder the equations read
\begin{equation}
 \tau \frac{da_n}{dt}=-\gamma a_n-i \kappa  \left\{ \exp(-h) a_{n+1}+\exp(h) a_{n-1} \right\}
\end{equation}
 where $\tau$ is the round-trip time in the main rings, $\gamma$ their single-pass loss rate, and $\kappa$ is an effective coupling rate, defined by Eq.(26) given in the Methods. Note that, after the substitution $a_n(t)=c_n(t) \exp(-\gamma t / \tau)$, Eqs.(9) reduces to the Hatano-Nelson model (5) without disorder. It should be noted that, if the circulation of the light fields in the rings is reversed, Eq.(9) still holds after the substitution $h \rightarrow -h$, i.e. the direction of robust transport is reversed. However, provided that light circulation in either one of the two directions is not excited or prevented, the CROW structure realizes the non-reciprocal  Hatano-Nelson model (5). In the CROW system, the disorder $V_n$ arises because of deviations of microring resonance frequencies from the reference value due to fabrication imperfections. Deviations of the antiresonance condition in the auxiliary rings also contribute to the disorder $V_n$ (see Methods).
 In the Hermitian limit $h=0$, i.e. in the absence of the gain/loss regions in the auxiliary rings, disorder is responsible for Anderson localization, which has been experimentally observed in Ref.\cite{r25}. The introduction of the imaginary gauge field should prevent Anderson localization. \par 
 Here we suggest a simple experiment for the observation of the non-Hermitian delocalization transition that occurs in the presence of the imaginary magnetic field. To this aim, let us consider a finite CROW chain  
 made of $N \gg 1$ main microrings, in which the ring in one edge of the chain (for instance the ring  with index $n=1$) is pumped and provides a gain parameter $g_0$ per round-trip; see Fig.6(a). Taking into account disorder of the microring resonance frequencies and the gain in the edge microring, coupled-mode equations are modified as follows
 \begin{equation}
 \tau \frac{da_n}{dt}=(-\gamma + g_0 \delta_{n,1}-iV_n) a_n-i \kappa  \left\{ \exp(-h) a_{n+1}+\exp(h) a_{n-1} \right\}
 \end{equation}
 $(n=1,2,3,...,N$). The CROW structure sustains a set of $N$ modes (also called supermodes), which are obtained as the eigenvectors of the linear system (10). 
 For a given realization of disorder $V_n$ and for $g_0=0$, all modes are damped in time, because their associated eigenvalues have a negative real part. As the gain parameter $g_0$ is increased up to some threshold value, a neutrally stable (i.e. undamped) mode will arise among the $N$ supermodes, which corresponds to the lasing mode of the CROW structure. 
 The localization properties of the lasing mode turn out to be strongly sensitive on the value of $h$, i.e. of the imaginary gauge field. In fact, the $N$ supermodes sustained by the active microring chain and their corresponding thresholds $g_0$  are obtained from Eq.(10) after making the Ansatz 
 \begin{equation}
 a_n(t)=p_n \exp(i \lambda t / \tau ), 
\end{equation} 
  where $\lambda$ is a real parameter that defines the frequency detuning of the supermode from the microring reference frequency. Substitution of Eq.(11) into Eq.(10) yields
  \begin{equation}
( i \lambda +\gamma - g_0 \delta_{n,1})p_n=-iV_n p_n-i \kappa  \left\{ \exp(-h) p_{n+1}+\exp(h) p_{n-1} \right\}.
 \end{equation}
  If the gain ($g_0$) and loss ($\gamma$) parameters are smaller or of the same order than $\kappa$, the mode distribution $|p_n|^2$ is not much distorted from the one of the chain with $\gamma=g_0=0$. Hence the localization property of the lasing mode is sensitive to the absence or the presence of the imaginary gauge field $h$. For $h=0$ and for a sufficiently strong disorder, because of Anderson localization all the modes are localized in a few sites of the chain with a localization length $\xi <N$, and the lasing mode corresponds to the localized state $p_n(h=0)$ with the largest occupation at the pumped ring $n=1$ [Figs.6(b) and (c)]. For $h \neq 0$, the distribution of the lasing mode is simply given by $p_n(h)=p_n(h=0) \exp(hn)$, as it readily follows from Eq.(12) after the transformation $p_n \rightarrow p_n \exp(hn)$. Hence, as $h$ is increased such that $2h$ gets comparable to the inverse of the localization length $1/ \mathcal{\xi}$, the mode $p_n(h)$  is delocalized all along the chain or even localized closest to the opposite edge $n=N$ if $h$ is further increased. Hence the oscillating mode will be delocalized [Fig.6(d)]. In an experiment, the delocalization effect induced by the imaginary gauge field $h$ can be thus simply visualized as a delocalization transition of the lasing intensity distribution in the microring chain as the parameter $h$ is varied. Such a simple experiment might provide the first accessible testbed for the observation of the non-Hermitian delocalization transition predicted in the pioneering work by Hatano and Nelson.  
  
\vspace{0.3cm}

\section*{Discussion}
Non-Hermitian photonic lattices with asymmetric amplification/attenuation of counter-propagating modes can provide a route toward robust light transport, preventing Anderson localization in the presence of disorder. The physical origin of one-way transport in non-Hermitian lattices considered in our work is rooted in the non-Hermitian delocalization transition originally discussed in quantum mechanics by Hatano and Nelson for a lattice with an imaginary gauge potential \cite{r18}. It is thus very distinct than the physics of topologically-protected edge modes arising in 2D photonic lattices  with synthetic gauge fields \cite{r1}, or the physics of asymmetric transport in other $\mathcal{PT}$-symmetric systems where non-reciprocity is obtained by exploiting some nonlinearities in the system \cite{palle1,palle2} . We have suggested a very simple physical condition for  non-Hermitian robust one-way transport and non-Hermitian delocalization [see Eq.(1)],  and shown that the Hatano Nelson model provides a special example where such a condition is met. We conjectured the generality of such a condition, that can be satisfied in driven non-Hermitian lattices as well (see Fig.4). Finally, we have proposed a simple method for the implementation of an imaginary gauge field for photons in coupled microring resonators, which could thus provide the first experimentally accessible testbed for the observation of the Hatano-Nelson non-Hermitian delocalization. Our results indicate that  non-Hermitian photonic lattices can provide a new route toward unidirectional light transport which is robust to imperfections and disorder, and could be of relevance to other related phenomena in photonics, such as  non reciprocal photonic transmission, isolation, and unidirectional coherent perfect absorption \cite{r26,r27,r28,r29}, as well as in other non-Hermitian physical systems such as Josephson circulators \cite{r28,r29}.

\par

\section*{Methods}

{\bf Quasi-energy spectrum of the driven non-Hermitian lattice.} The quasi-energy spectrum of the Hamiltonian (7) in the absence of disorder ($V_n=0$) can be calculated as follows.
After setting $|\psi(t) \rangle= \sum_n a_n(t) \exp(-iFnt) | n \rangle$, the evolution equations for the amplitude probabilities $a_n$ read
\begin{equation}
i \frac{da_n}{dt}= \kappa \left\{  \exp(-iFt) a_{n+1} +\exp(iFt) a_{n-1} \right\}+(-1)^n g(t) a_n.
\end{equation}
We look for a solution to Eq.(13) of the form
\begin{equation}
a_n(t)= \left(
\begin{array}{c}
A(t) \\
B(t)
\end{array}
\right) \exp(-i q n/2)
\end{equation}
where the upper (lower) row applies to an even (odd) value of $n$, and where $q$ is the Bloch wave number that varies in the range $(-\pi,\pi)$. Substitution of Eq.(14) into Eq.(13) yields
\begin{eqnarray}
i \frac{dA}{dt} & = & 2 \kappa B \cos \left( q/2+Ft \right)+g(t)A \\
i \frac{dB}{dt} & = & 2 \kappa A \cos \left( q/2+Ft \right)-g(t)B. 
\end{eqnarray}
If the resonance condition $F=(M/N) \omega$ is satisfied, Eqs.(15) and (16) are periodic in time with period $T= 2 \pi N / \omega$ and Floquet theory applies. Indicating by $\Phi(q)=\exp( - iT \mathcal{R}(q))$ the propagator of Eqs.(15) and (16) from $t=0$ to $t=T$, i.e. $(A(T),B(T)^T=\Phi(q) (A(0),B(0))^T$, the quasi energies $E_{\pm}(q)$ of the two mini bands for the Hamiltonian $\hat{H}$  are defined as the eigenvalues of the $2 \times 2$ matrix $\mathcal{R}(q)$.  Quasi energies are defined apart from integer multiples than $\omega /N$; for the sake of definiteness the real part of the quasi energies is taken within the interval $(-\omega/2N, \omega/2N)$. Note that, since the modulation function $g(t)$ satisfies the condition $g(-t)=g^*(t)$, then it can be readily shown that $E_{\pm}(-q)=E_{\pm}^*(q)$.

\vspace{0.2cm}
{\bf Wave packet distortion effects.}  Let us consider an initially localized wave packet with carrier wave number $q_0$ that propagates in the ordered lattice. The initial excitation can be written rather generally as $c_n(0) \equiv c(n,0)=F(n) \exp(iq_0 n)=\int_{-\pi}^{\pi} dq S(q) \exp(iqn)$ with a slowly-varying amplitude $F(n)$, where $S(q)$ is the Bloch spectrum which is assumed to be a narrow function at around $q=q_0$.  For a static lattice with a single tight-binding band, like for the Hatano-Nelson model (2), the amplitude probabilities $c_n(t)$ evolve according to $c(n,t) = \int_{-\pi}^{\pi} dq S(q)  \exp [iqn-i E(q)t]= \int_{-\pi}^{\pi} dq S(q) \exp[{\rm Im} E(q)t] \exp[iqn-i {\rm ReE}(q)t]$, where $E(q)$ is the complex energy dispersion curve of the lattice band.  A similar expression holds for a time-periodic Hamiltonian [e.g. the model (7)], provided that time $t$ is taken at integer multiplies than the modulation period $ 2 \pi / \omega$ and $E(q)$ is replaced by the quasi-energy band. Owing to the dependence of the complex energy $E(q)$ on the Bloch wave number $q$, the wave packet is generally distorted during propagation. However, for a wave packet spectrally narrow at around the carrier wave number $q=q_0$ such that ${\rm Im} (dE / dq)_{q_0}=0$ and ${\rm Re} (d^2E / dq^2)_{q_0}=0$, at leading  order one has $|c(n,t)|^2 \simeq \exp(2gt) |c(n-v_gt,0)|^2$, where $g \equiv {\rm Im} E(q_0)$, and $v_g={\rm Re} (d E/dq)_{q_0}$. This means that the wave packet propagates nearly undistorted with a group velocity $v_g$, apart from a uniform amplification ($g>0$) or attenuation ($g<0$). In the Hatano-Nelson Hamiltonian the condition of nearly-undistorted propagation is attained at $q_0=\pi/2$, whereas for the driven lattice model (7) nearly undistorted propagation is predicted for almost any wave number $q_0$, owing to the flatness of ${\rm Im}E(q)$ and the linear dependence of ${\rm Re} E(q)$ on $q$ inside the Brillouin zone [Fig.3(a)].   As an example, Fig.7(a) shows the numerically-computed evolution of a two-humped wave packet $F(n)=n \exp[-(n/10)^2]$ for the Hatano-Nelson Hamiltonian for carrier wave number $q_0=0$ and $q_0= \pi/2$, in either an ordered and a disordered lattice. The figure clearly shows that, according to the previous analysis, even in the absence of disorder strong wave packet distortion is observed for $q_0=0$, whereas wave packet distortion is much weaker for $q_0= \pi/2$. For the modulated lattice Hamiltonian [Eq.(7)], wave packet  distortion is almost absent, as shown in Fig.7(b). The reason thereof is the special behavior of dispersion curves of the quasi energy minibands of the driven lattice, as discussed in the main text.

\vspace{0.2cm}
{\bf Imaginary gauge field in a chain of microrings: coupled-mode analysis.}  
For the sake of clearness, let us consider the case of two high-$Q$ rings in the main chain coupled by an auxialiay ring [Fig.4(b)]. The method can be readily extended to an arbitrary number of rings in the main cavity indirectly coupled by auxiliary rings.  The fields in the two rings of the main chain are assumed to propagate counterclockwise, whereas the field in the auxiliary ring propagates clockwise. The carrier frequency of the fields is assumed to be one longitudinal frequency of the rings in the main chain.
Indicating by $a(t)$, $a'(t)$, $b(t)$, $b'(t)$, $e(t)$, $f(t)$, $g(t)$ and $l(t)$ the field amplitudes in the main and auxiliary rings at the locations depicted in Fig.4(b), resonator coupling is described by the following equations \cite{r24}
\begin{equation}
\left(
\begin{array}{c}
a' \\
f
\end{array}
\right)
=
\left(
\begin{array}{cc}
u & \rho \\
-\rho^* & u
\end{array}
\right)
\left(
\begin{array}{c}
a \\
e
\end{array}
\right) \; , \; \;\;
\left(
\begin{array}{c}
b' \\
l
\end{array}
\right)
=
\left(
\begin{array}{cc}
u & \rho \\
-\rho^* & u
\end{array}
\right)
\left(
\begin{array}{c}
b \\
g
\end{array}
\right)
\end{equation}
where $\rho$ is the coupling constant and $|u|^2+|\rho|^2=1$. We assume small coupling, i.e. $|\rho| \ll 1$ and $u \simeq 1$. A real value of $u$ is also assumed for the sake of clearness.
The field boundary conditions for the two main rings read
\begin{equation}
a(t+\tau)=a'(t) \exp(-\gamma) \simeq (1-\gamma) a'(t)\; \; ,\;\;\; b(t+\tau)=b'(t) \exp(-\gamma)  \simeq (1-\gamma) b'(t)
\end{equation}
where $\gamma \ll 1$ is the single-pass loss in each ring and $\tau$ is the transit time in the ring. Taking into account propagative effects in the two half sections of the auxiliary ring one can also write
\begin{equation}
g(t+\tau_1)=f(t)\exp(h+i \phi_1) \; , \;\; e(t+\tau_1)=l(t)\exp(-h+i \phi_2)
\end{equation}
where $2 \tau_1 \simeq \tau$ is the propagation time in the auxiliary ring, $\phi_{1,2}$ are the accumulated phases due to field propagation in the upper and lower sections of the auxiliary ring, and $h$ is the gain/loss parameter in each of the two sections. Since the auxiliary ring is assumed to be in anti-resonance, one has $\phi_1+\phi_2=\pi$ (apart from integer multiplies than $2 \pi$). The anti-resonance condition is obtained by slightly shortening (or lengthening) the length of the auxiliary ring as compared to the rings in the main chain. In the single longitudinal mode regime and assuming that the field amplitudes vary slowly over one round trip in the rings (mean-field limit), i.e. $a(t+ \tau) \simeq a(t)+(da/dt) \tau$, $b(t+ \tau) \simeq b(t)+(db/dt) \tau$, etc. from Eqs.(17) and (18) one can write
\begin{equation}
\tau \frac{da}{dt} \simeq -(1-u)a-\gamma a + \rho e \;,\;\;\; \tau \frac{db}{dt} \simeq -(1-u)b-\gamma b + \rho g.
\end{equation}
On the other hand, from Eqs.(17-19) one has
\begin{eqnarray}
g(t+ \tau_1)+u^2 g(t-\tau_1) & = & u \rho^* b(t-\tau_1)- \rho^* a(t) \exp(h+i \phi_1) \\
e(t+\tau_1)+u^2 e(t-\tau_1) & = &  u \rho^* a(t-\tau_1)-\rho^* b(t) \exp(-h+i \phi_2)
\end{eqnarray}
where we used the anti-resonance condition $\phi_1+\phi_2= \pi$. For $|\rho| \ll1$, $u \simeq 1$ at leading order the delay effects in the amplitudes $g$ and $e$ can be neglected, thus obtaining
\begin{equation}
g(t) \simeq \frac{\rho^*}{2} \left[ u b(t)- a(t) \exp(h+i \phi_1) \right] \; ,\;\;\; e(t) \simeq \frac{\rho^*}{2} \left[ -b(t) \exp(-h+i \phi_2)+u a (t) \right].
\end{equation}
Assuming that the two gain/loss sections in the auxiliary rings have the same length, i.e. $\phi_1=\phi_2= \pi/2$, one obtains
\begin{equation}
g(t) \simeq \frac{\rho^*}{2} \left[ u b(t)- i a(t) \exp(h) \right] \; ,\;\;\; e(t) \simeq \frac{\rho^*}{2} \left[ -i b(t) \exp(-h)+u a (t) \right].
\end{equation}
Substitution of Eqs.(24) into Eqs.(20)  yields
\begin{equation}
\tau \frac{da}{dt}= - \left( 1-u-u | \rho|^2/2 \right)a-\gamma a-i \kappa \exp(-h) b \; ,\;\;\; \tau \frac{db}{dt}= - \left( 1-u-u | \rho|^2/2 \right)b-\gamma b-i \kappa \exp(h) a 
\end{equation}
 where we have set
 \begin{equation}
 \kappa \equiv \frac{|\rho|^2}{2}.
 \end{equation}
Taking into account that $u=\sqrt{1-|\rho|^2} \simeq 1- |\rho|^2/2$, at leading order from Eqs.(25) one finally obtains
\begin{equation}
\tau \frac{da}{dt}=-\gamma a-i \kappa \exp(-h) b \; ,\;\;\; \tau \frac{db}{dt}= -\gamma b-i \kappa \exp(h) a.
\end{equation}
Such equations clearly show that the role of the auxiliary ring is to indirectly couple the two main rings, with unbalances hopping rates $\kappa \exp(-h)$ and $\kappa \exp(h)$, where $h$ is the gain/loss parameter in the two half-sections of the auxiliary ring and $\kappa$ is defined by Eq.(26). \par In the above analysis we assumed exact anti-resonance of the auxiliary resonator, i.e. $\phi_1=\phi_2= \pi/2$. A slight deviation of the auxiliary ring from the anti-resonance condition can be taken into account by assuming in the previous analysis $\phi_1=\pi/2+ \omega_1$ and $\phi_2= \pi/2+ \omega_2$, where $\omega_{1,2}$ are additional phase deviations from the ideal condition that could arise in fabrication imperfections. Anti-resonance is broken whenever $\omega_1+\omega_2 \neq 0$. For small values of  phase deviations $\omega_{1}$ and $\omega_2$, after some straightforward calculations it can be shown that at leading order Eq.(27) is modified as follows
\begin{equation}
\tau \frac{da}{dt}=\left( -\gamma +i \kappa \frac{\omega_1+\omega_2}{2} \right) a-i \kappa \exp(-h+i \theta) b \; ,\;\;\; \tau \frac{db}{dt}=\left( -\gamma +i \kappa \frac{\omega_1+\omega_2}{2} \right) b-i \kappa \exp(h-i \theta) a,
\end{equation}
where we have set $\theta \equiv (\omega_2-\omega_1)/2$. Equation (28) shows that deviation from the anti-resonance condition has two effects: it introduces an effective shift of the resonance frequency of the main rings by the amount $\kappa( \omega_1+ \omega_2)/(2 \tau)$, and it provides a phase term $\theta$ for the hoping rate $\kappa$. The latter does not play any role since it can be removed by a proper gauge transformation. Therefore, slight deviations from the anti-resonance condition of the auxiliary ring can be accounted for by a disorder $V_n$ of the main ring resonance frequencies, like in Eq.(10). 

\bibliography{MEP_BIB}

{\small 
\acknowledgements{
This work was partially supported by the Fondazione Cariplo (Grant
No. 2011-0338).}

\section*{Author contributions}
SL conceived the idea, SL and GDV developed the theory, SL, DG and GDV made the numerical simulations. All authors discussed the results and participated in the manuscript preparation.


\section*{Competing financial interests}
The authors declare no competing financial interests.

\newpage

\section*{Figure Captions}
\vspace{1cm}
\noindent
{\bf Fig.1} Principle of robust light transport based on (a) topologically-protected edge states, and (b) asymmetric non-Hermitian transport.\\
\\
{\bf Fig.2} Numerically-computed evolution of the wave packet center of mass $\langle n(t) \rangle$ in a semi-infinite Hermitian lattice described by the Hamiltonian (4) corresponding to single-site excitation. (a) Edge site excitation [$c_n(0)=\delta_{n,0}$] in the Hermitian lattice ($h=0$), (b) edge-site excitation  [$c_n(0)=\delta_{n,0}$] of the the Non-Hermitian lattice ($h=0.2$), (c) site excitation in the bulk  [$c_n(0)=\delta_{n,100}$] of the the Non-Hermitian lattice ($h=0.2$). The dashed curves show the center of mass path in the absence of disorder, whereas the solid thin curves show the paths corresponding to a disordered lattice for 16 realizations of disorder. $V_n$ is assumed to have a uniform distribution in the range $(-\kappa, \kappa)$. The lower panels in (a) and (b) depict the distribution of the occupation probabilities $|"c_n(t)|^2$ (in arbitrary units) at time $t=40 / \kappa$ for the ordered lattice and for one realization of the disordered lattice. The lower panel in (c) show the evolution of the normalized occupation amplitude $a_n(t)$ in a pseudo color map (modulus square of $a_n(t)$). \\  
\\
{\bf Fig.3}  Wave packet evolution in a tight-binding lattice with two potential defects $V_0$ at sites $n_0$ and $n_1$ (upper inset) for (a) the Hermitian lattice ($h=0$), and (b) in the presence of an imaginary gauge field ($h=0.25$). Parameter values are $V_0 / \kappa =1$ and $n_1-n_0=20$. Initial condition corresponds to the Gaussian wave packet $c_n(0) \propto \exp[-(n-100)^2/64+i \pi n/2]$. The figures show the evolution of the normalized amplitude probabilities $a_n(t)$ (modulus of $a_n(t)$) is a pseudo color map; the two vertical dashed lines in the maps show the position of the potential defects. 
While in the Hermitian case the two potential defects cause multiple wave packet reflection back and forth and echoes of the transmitted wave packet, multiple reflections and echoes are suppressed when the imaginary magnetic field is applied.\\
\\
{\bf Fig.4}  (a) Numerically-computed quasi-energy spectrum (real and imaginary parts of the quasi energies $E_{\pm}$) of the non-Hermitian driven Hamiltonian (7) in the absence of disorder for parameter values $\omega / \kappa =1$, $F/ \omega=2$, $G_R / \kappa=4.7$ and $G_I / \kappa=4.26$. The filled dotted curves correspond to forward propagating modes ($v_g= {\rm Re}(dE/dq)>0$) and positive imaginary quasi energy, whereas the open dotted curves correspond to the backward propagating waves. 
 The dashed curve in the upper plot shows, for comparison, the quasi energy spectrum in the Hermitian case $G_I=0$. (b) Numerically-computed evolution of the wave packet center of mass $\langle n(t) \rangle$ in the semi-infinite Hermitian lattice corresponding to initial excitation of the edge site. The dashed curve shows the path in the absence of disorder, whereas the solid thin curves show the paths corresponding to a disordered lattice for a a few realizations of disorder. $V_n$ is assumed to have a uniform distribution in the range $(-\kappa, \kappa)$. (c) Same as (b), but for the non-Hermitian lattice.\\
\\
{\bf Fig.5} Optical realization of an imaginary gauge field in coupled microresonators. (a) Schematic of the CROW structure, comprising a sequence of main microrings indirectly coupled via auxiliary rings. The auxiliary rings provide light amplification (attenuation) in the upper (lower) half perimeter of the microring. (b) Schematic of indirect coupling of two main microrings via an auxiliary microring.\\
\\
{\bf Fig.6} (a) Schematic of a CROW laser made of $N$ coupled microrings. (b) Distribution of resonance frequency detunings arising from disorder in a chain made of $N=60$ microrings (behavior of $V_n / \kappa$). (c),(d) Intensity distribution (in arbitrary units) of the lowest threshold lasing mode in the Hermitian ($h=0$, panel c), and non-Hermitian ($h=0.05$, panel d) cases.\\
\\
{\bf Fig.7} Numerically-computed  wave packet evolution in (a) the Hatano-Nelson Hamiltonian [Eq.(4)], and (b) in the driven lattice Hamiltonian [Eq.(8)] in the absence (left panels) and in the presence (right panels) of disorder. The strength of disorder is as in Figs.2 and 4. The figures show the evolution of the normalized amplitude probabilities $a_n(t)$ (modulus of $a_n(t)$) is a pseudo color map. Initial condition corresponds to a two-humped wave packet $c_n(0) \propto n \exp[-(n/10)^2+i q_0 n]$ with carrier Bloch wave number $q_0=\pi /2$ (upper plots) and $q_0=0$ (lower plots). \\

\newpage
{\bf Fig.1}
\includegraphics[width=16cm]{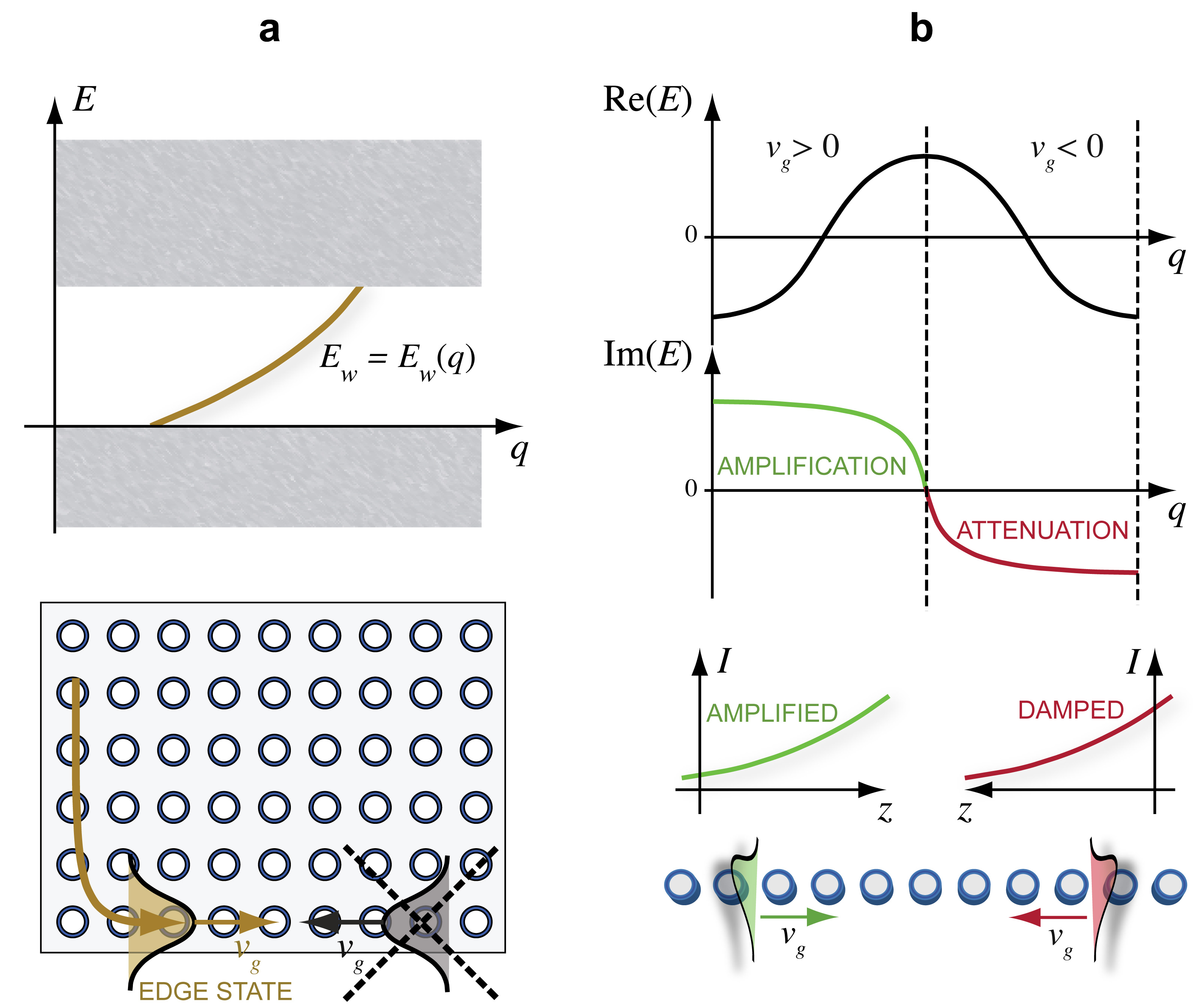}

\newpage
{\bf Fig.2}
\includegraphics[width=16cm]{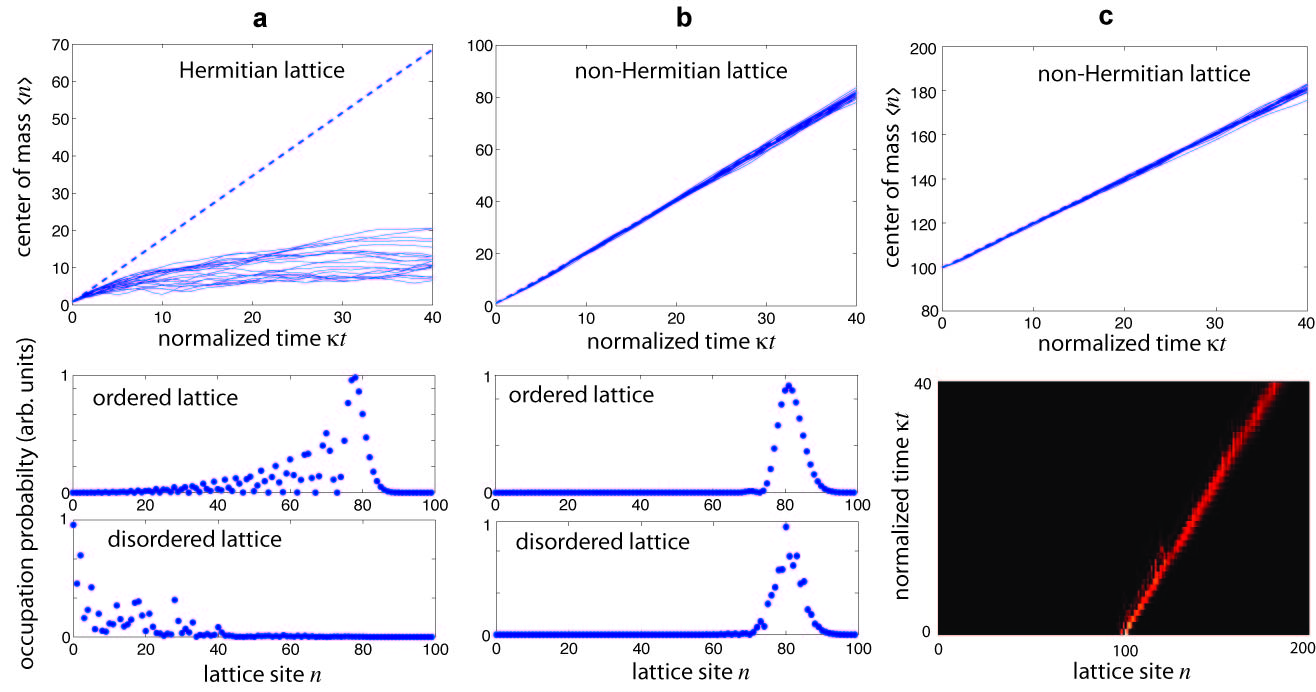}

\newpage
{\bf Fig.3}
\includegraphics[width=25cm]{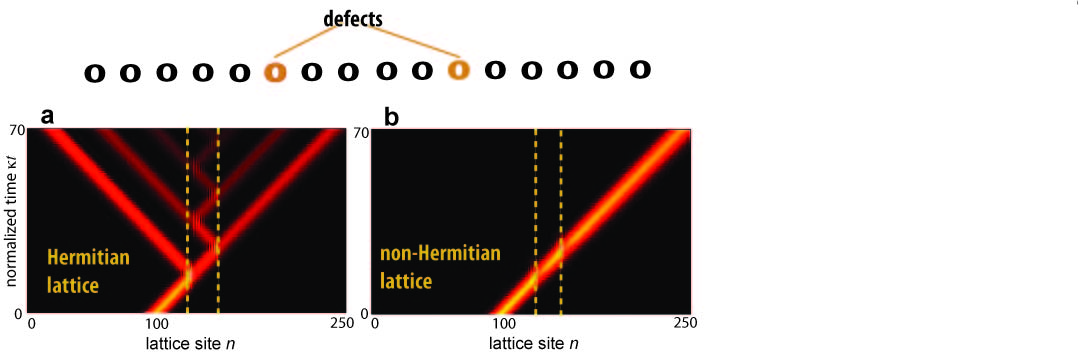}

\newpage
{\bf Fig.4}
\includegraphics[width=18cm]{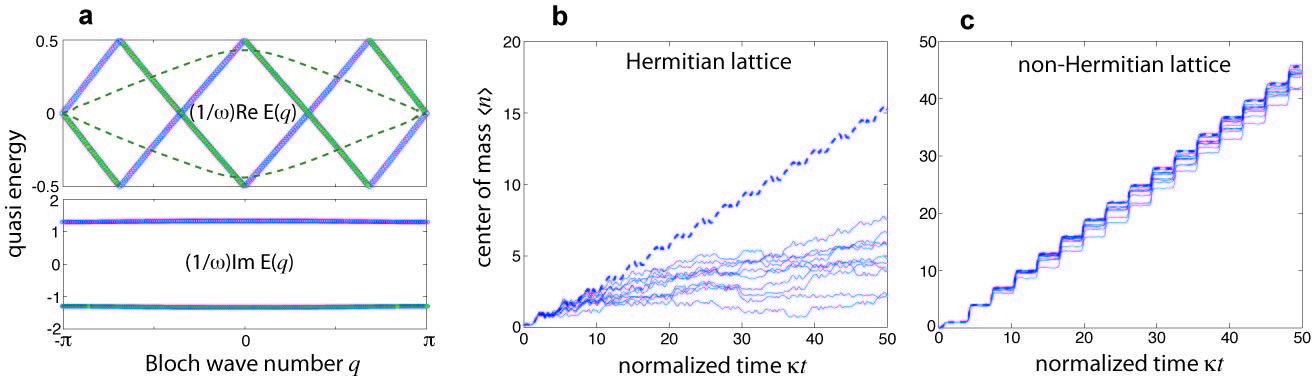}

\newpage
{\bf Fig.5}
\includegraphics[width=16cm]{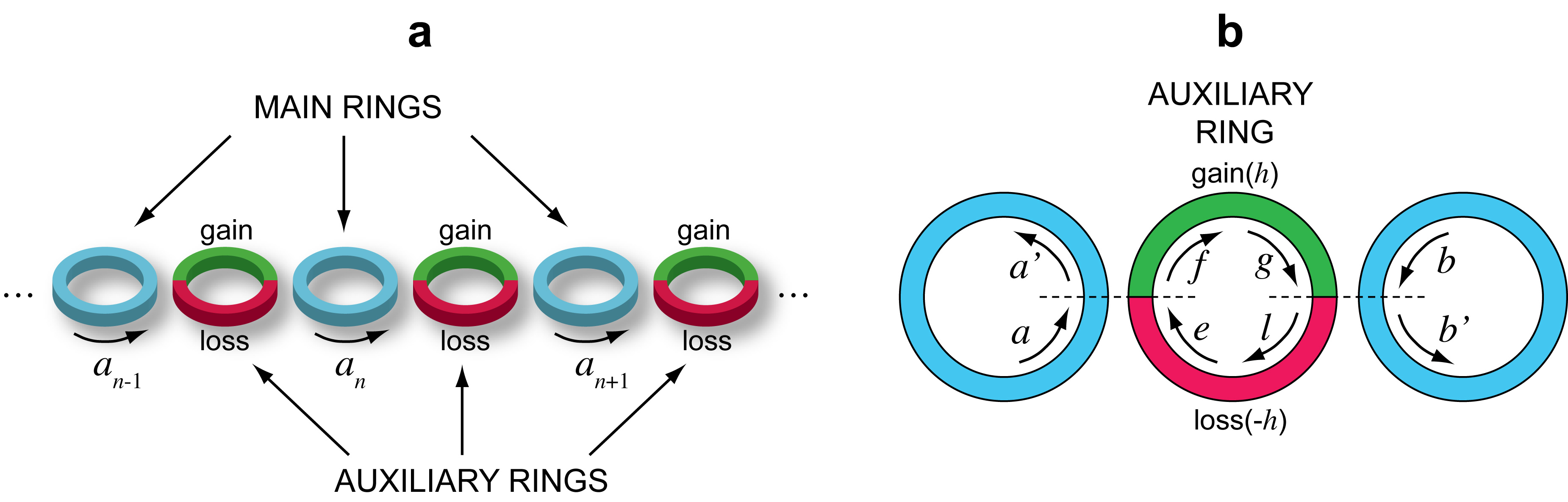}

\newpage
{\bf Fig.6}
\includegraphics[width=16cm]{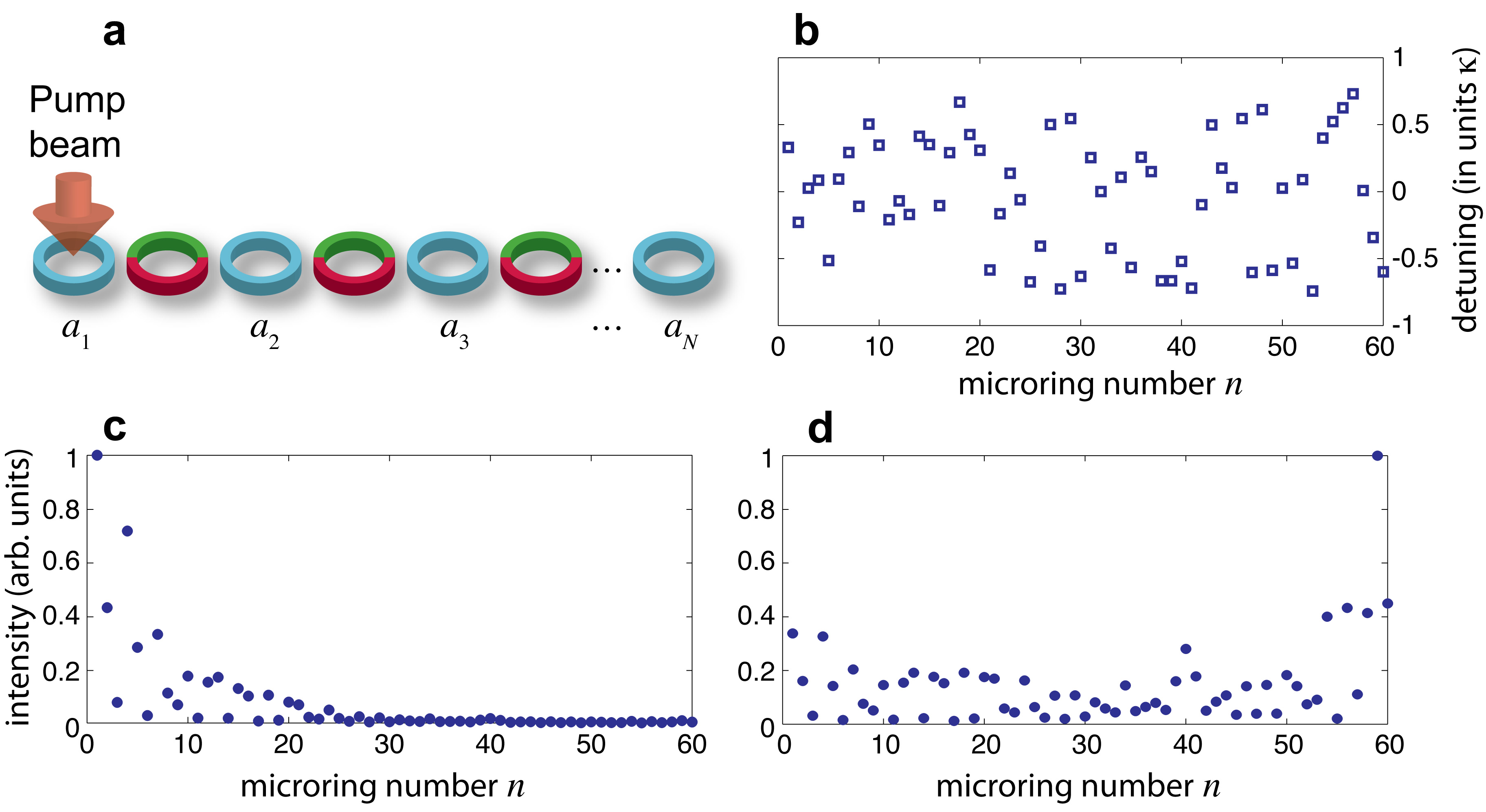}

\newpage
{\bf Fig.7}
\includegraphics[width=16cm]{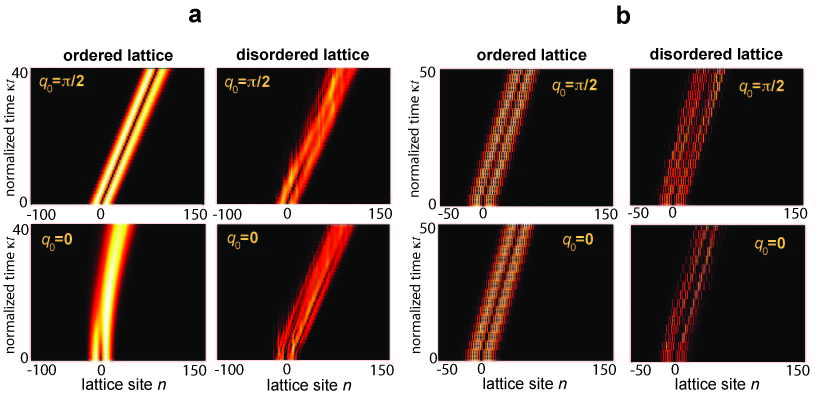}

\end{document}